# Spectral correlations in disordered electronic systems: Crossover from metal to insulator regime

Arkady G. Aronov[1,2], Vladimir E. Kravtsov[2,3], and Igor V. Lerner[4]

[1] *Weizmann Institute of Science, Department of Condensed Matter Physics, 76100 Rehovot, Israel*
[2] *International Centre for Theoretical Physics, 34100 Trieste, Italy*
[3] *Institute of Spectroscopy, Russian Academy of Sciences, 142092 Troitsk, Moscow r-n, Russia*
[4] *School of Physics and Space Research, University of Birmingham, Birmingham B15 2TT, United Kingdom*

We use the semiclassical approach combined with the scaling results for the diffusion coefficient to consider the two-level correlation function $R(\varepsilon)$ for a disordered electron system in the crossover region, characterized by the appearance of a macroscopic correlation or localization length, $\xi$, that diverges at the metal-insulator transition. We show new critical statistics, characterized by a non-trivial asymptotic behavior of $R(\varepsilon)$, to emerge on both sides of the transition at higher energies, and to expand to all energies larger than mean level spacing when $\xi$ exceeds the system size.

Single-electron spectra in disordered metals are governed [1–3] by the universal Wigner–Dyson (WD) statistics [4] which are applicable to a large variety of different quantum systems. The principal conjecture is that spectra of complicated systems are statistically equivalent to the eigenvalues of the random matrix hamiltonians [5] constrained due to the presence of some general symmetries. The main feature of the WD statistics is the level repulsion at all energy scales. For a disordered system, it is due to the electronic eigenstates being extended.

With the increase of disorder, the system undergoes the Anderson metal–insulator transition [6]. At the other side of the transition, in the insulating phase, there is no correlation between the energy levels of the localized states, and the spectral statistics prove to be Poisson.

As it has become clear recently, there should exist the third universal level statistics, applicable exactly at the transition point. Its existence has been first asserted by Shklovskii et al. [7] who have suggested that the nearest-level probability distribution, $P(\varepsilon)$, was a universal hybrid of the WD one at small $\varepsilon$ and the Poisson one at large $\varepsilon$. Systematic analytical studies of the universal critical statistics have been started in Ref. [8]. Namely, the two-level correlation function, has been studied in the framework of the diagrammatic approach and the scaling hypothesis, and found to be drastically different from both known universal statistics. In contrast to the Poisson statistics, the long-range level correlations are still present, but they are different from the WD ones. The mapping of the analytical results of Ref. [8] onto the one-dimensional plasma model with the power-law interaction [9] illustrated the presence of the level repulsion at large energies. It shows itself in $P(\varepsilon)$ having the large energy asymptotic behavior intermediate between Gaussian for the WD statistics and exponential for the Poisson ones [9], rather than being just Poisson as in Ref. [7].

All the three universal statistics are exactly applicable in the thermodynamic limit, $L_0 \to \infty$. The critical statistics are applicable within an energy band with a fixed (but arbitrary large) number of the levels, centered exactly at the mobility edge, $\varepsilon_{\scriptscriptstyle 0} = \varepsilon_c$, while the WD and Poisson statistics are applicable for energy bands with $\varepsilon_{\scriptscriptstyle 0} > \varepsilon_c$ (the metallic region) and $\varepsilon_{\scriptscriptstyle 0} < \varepsilon_c$ (the insulating region), respectively. However, when the sample size $L_0$ is finite, there is a smooth crossover between the metallic and the insulating phase. In this Letter, we present the analytical description of an appropriate crossover between the WD and Poisson statistics.

The crossover region is characterized by the appearance, on approaching the Anderson transition, of a new macroscopic scale, $\xi$, which is the correlation or localization length at the metallic or insulating side of the transition, respectively. It diverges at the mobility edge as [6]

$$\xi = \ell_{\scriptscriptstyle 0}\, |1 - g_{\scriptscriptstyle 0}/g_c|^{-\nu}, \qquad (1)$$

where $\ell_{\scriptscriptstyle 0}$ is some microscopic length (that can be of the order of the elastic scattering length, $\ell = v_F \tau$), $g_{\scriptscriptstyle 0}$ is the conductance at the scale $\ell_{\scriptscriptstyle 0}$, $g_c$ is its critical value, and $\nu$ is the critical exponent that depends on the dimensionality, $d$, and the universality class. We will show the new critical statistics to emerge at large energies in the crossover region, when $\ell_{\scriptscriptstyle 0} \ll \xi < L_0$, and to become universal and applicable to all energies [10] $\varepsilon \gg \Delta$ (with $\Delta$ being the mean level spacing) near the transition, when $\xi \gtrsim L_0$.

Calculations in the crossover region may be performed rigorously within the diagrammatic approach, similar to that used at the mobility edge [8]. In this Letter we will use instead a simplified description based on a more intuitive and illuminating way of treating the spectral correlations developed recently by Argaman, Imry, and Smilansky [11] within the semiclassical approach.

We consider the two-level correlation function (TLCF):

$$R(s) = \frac{1}{\rho^2} \left\langle \rho\!\left(E - \frac{\varepsilon}{2}\right) \rho\!\left(E + \frac{\varepsilon}{2}\right) \right\rangle - 1, \quad s \equiv \frac{\varepsilon}{\Delta}. \qquad (2)$$

Here $\rho(E)$ is the electron density of states at the energy $E$ for a particular realization of disorder, $\langle \ldots \rangle$ stand for averaging over all the realizations, $\rho \equiv \langle \rho(E) \rangle$, and $\Delta = 1/(\rho L_0^d)$. The spectral form factor is defined as

$$K(t) = \int \frac{d\varepsilon}{2\pi\hbar} R(\varepsilon) e^{-i\varepsilon t/\hbar} . \qquad (3)$$

A transparent semiclassical expression for it, based on the Gutzwiller trace formula [12], has been obtained by Berry [13]. Combining the Berry's expression with a generalized Hannay and Ozorio de Almeida's sum rule [14], Argaman, Imry, and Smilansky [11] have shown that

$$K(t) \simeq \frac{2|t|\Delta}{(2\pi\hbar)^2 \beta} p(t), \qquad (4)$$

where $\beta = 1, 2,$ or $4$ is for the Dyson unitary, orthogonal, and symplectic ensembles, respectively, $p(t)$ is the averaged classical probability density to perform periodic motion of period $t$ at a *given energy* [15]. It is different from the averaged classical return probability for a diffusing electron (which is contributed to by any path returning the point of origin, including those with mismatching initial and final momenta) simply by a constant factor.

One identifies two opposite time scales: ergodic, required for the diffusion motion to fill the entire phase space, $t \gg \tau_{\rm erg} \equiv L_0^2/D$, and diffusive, $t \lesssim \tau_{\rm erg}$, with $D$ being the diffusion coefficient. (We involve here neither ballistic, $t \lesssim \tau$, nor quantum, $t \gtrsim \hbar/\Delta$, regimes). They correspond to the energy scales $s \ll g$ and $s \gtrsim g$, respectively, where $g = E_c/\Delta = \hbar \rho D L_0^{d-2}$ is the conductance in units of $e^2/\hbar$, and $E_c = \hbar/\tau_{\rm erg}$ is the Thouless energy.

Efetov has shown [2] the statistics in the ergodic regime to be the same as in the Wigner–Dyson random matrix theory [5]. There is no dependence either on $d$, or on $D$, as only the spatially homogeneous diffusion mode (i.e. the $q = 0$ Fourier component of the diffusion propagator) contributes to the correlation function. It corresponds to the saturation of the probability density [13,11] in Eq. (4), $p(t)=$ const. Then $K(t) \propto t$ results in $R_{\rm erg}(s) \sim 1/s^2$, and thus in a logarithmic spectral rigidity [16].

At the shorter times, $t \lesssim \tau_{\rm erg}$, the probability distribution of a diffusing electron is a standard Gaussian:

$$P(\bm{r}, t) = \frac{1}{\rho (4\pi Dt)^{d/2}} \exp\left[-\frac{(\bm{r} - \bm{r}_{\rm o})^2}{4Dt}\right] . \qquad (5)$$

The return probability $p(t)$ is obtained by setting $\bm{r} = \bm{r}_{\rm o}$. Substituting it into Eq. (4), one obtains in the diffusive regime [11,15]:

$$K(t) \simeq \frac{2\Delta}{(2\pi\hbar)^2 \beta \rho (4\pi D)^{d/2}} t^{1-d/2} . \qquad (6)$$

Performing the Fourier transform, one reproduces the diagrammatic result of Altshuler and Shklovskii [3]:

$$R_{\rm dif}(s) = C_d g^{-d/2} s^{-(2-d/2)}, \qquad (7)$$

where $C_d$ is a positive constant for $d > 2$, $C_3 \sim 1$. Although $R_{\rm dif}(s)$ is not universal, it depends only on the conductance $g$ and the dimensionality $d$ so that the diffusive regime is governed by the one-parameter scaling. Note that it corresponds to the *levels attraction* [17].

We can still use such a semiclassical approach to describe the spectral statistics in the crossover region and even at the mobility edge. The trick is to substitute the diffusion constant in Eq. (6) by the scale-dependent diffusion coefficient known from the scaling theory of localization [6].

In the crossover region, $\ell_{\rm o} \ll \xi \lesssim L_0$, the natural energy scale related to $\xi$, Eq. (1), is

$$\Delta_\xi = \frac{1}{\rho \xi^d} \equiv \left(\frac{L_0}{\xi}\right)^d \Delta, \qquad (8)$$

that is the mean level spacing within a correlation volume. It is known from the scaling theory of localization [6] that at scale $L \gtrsim \xi$ the conductance shows an Ohmic behavior, $g = \sigma L^{d-2}$, with $\sigma = g_c \xi^{2-d}$. The electron propagation at this scale remains diffusive, with the scale-independent diffusion constant

$$D_\xi \simeq \frac{g_c}{\hbar \rho \xi^{d-2}} = \frac{g_c}{g_{\rm o}} \left(\frac{\ell_{\rm o}}{\xi}\right)^{d-2} D \simeq \left(\frac{\ell_{\rm o}}{\xi}\right)^{d-2} D, \quad L \gtrsim \xi. \qquad (9)$$

The corresponding time scale is given by $t \gtrsim \xi^2/D_\xi \simeq \hbar/g_c \Delta_\xi \sim \hbar/\Delta_\xi$, as $g_c \sim 1$ for $d = 3$. At this time scale, the TLCF is still given by Eq. (7) as the rescaling of $D$ is absorbed by an appropriate rescaling of $g$.

Totally new statistics emerge at the larger energies [10], in the *critical regime*,

$$s \gtrsim \Delta_\xi/\Delta \quad \Longleftrightarrow \quad t \lesssim \hbar/\Delta_\xi, \qquad (10)$$

corresponding to the diffusion at the scale $\Lambda \lesssim \xi$. In this regime, the conductance is almost scale-independent and the diffusion is anomalous, with the coefficient given by

$$D_{\rm cr}(\Lambda) = \frac{g_{\rm cr}(\Lambda)}{\hbar \rho \Lambda^{d-2}} . \qquad (11)$$

*To the first approximation*, $g_{\rm cr}(\Lambda) = g_c$, and the time dependence of the anomalous diffusion may be found by combining Eq. (11) and $\Lambda^2 = D_{\rm cr}(\Lambda) t$ which leads to the well-known [18] result:

$$D_{\rm cr}(t) \approx D_0(t) = (g_c/\hbar\rho)^{2/d} t^{-1+2/d} . \qquad (12)$$

Substituting this to Eq. (6), we find the form factor to be time-independent that results in

$$R_0(s) = 0, \qquad s \gtrsim \Delta_\xi/\Delta . \qquad (13)$$

Such an absence of the correlations in the critical regime of Eq. (10) results from using the self-consistent approximation (12). For $\Lambda \sim L_0 \ll \xi$, it corresponds to the absence of correlations at the mobility edge found in the same approximation diagrammatically [8]. (To prove the diagrammatic cancellation, it was necessary to use not only the scaling relations (12) but also the causality requirements which are quite subtle in the diagrammatic technique but automatically taken into account in the present formalism). It means that the self-consistent approach is insufficient to describe the spectral statistics in the critical regime. One must find $D_{cr}(\Lambda)$, Eq. (11), more accurately, allowing for the scaling dependence of $g_{cr}(\Lambda)$.

The standard scaling equation for the conductance, linearized near the critical point $g = g_c$, has the form [6]

$$\frac{d\ln g}{d\ln \Lambda} = \frac{1}{\nu}\frac{g-g_c}{g_c}. \quad (14)$$

Integrating this on the metallic side with the initial condition $g(\Lambda = \ell_o) = g_o$, one obtains in the critical regime

$$g_{cr}(\Lambda) = g_c\left[1 + (\Lambda/\xi)^{\frac{1}{\nu}}\right], \quad \Lambda \ll \xi \lesssim L_0, \quad (15)$$

with $\xi$ defined by Eq. (1). It yields, after substituting into Eq. (11), an additional time-dependence of the diffusion coefficient, on top of that in Eq. (12),

$$D_{cr}(t) = D_0(t)\left[1 + (g_c t \Delta_\xi/\hbar)^{\frac{1}{\nu d}}\right], \quad (16)$$

thus leading to the *time-dependent* form factor,

$$K_{cr}(t) = \frac{\Delta}{2^{1+d}\pi^{2+d/2}\beta\hbar g_c}\left[1 - \alpha\left(\frac{t\Delta_\xi}{\hbar}\right)^{\frac{1}{\nu d}}\right], \quad (17)$$

where $\alpha$ is a numerical coefficient of order 1. This time-dependence leads to the following result for the TLCF in the critical regime of Eq. (10):

$$R_{cr}(s) = -A_{d\beta}\left(\frac{\Delta_\xi}{\Delta}\right)^{\frac{1}{\nu d}}\left(\frac{1}{s}\right)^{1+\frac{1}{\nu d}}, \quad (18)$$

with $A_{d\beta}$ being a numerical coefficient depending only on the dimensionality and the universality class. Therefore, there are three regimes in the crossover region: ergodic for $t \gtrsim \tau_{\text{erg}}$ (including the quantum times, $t \gtrsim \hbar/\Delta$) where statistics are Wigner–Dyson, diffusive for $\tau_{\text{erg}} \gtrsim t \gtrsim \hbar/\Delta_\xi$ where statistics are the same as in the non-ergodic regime in metals, Eq. (7), and critical, Eq. (10), where the new statistics emerge [19] described by the TLCF (18). When $\xi$ increases in approaching the transition, Eq. (1), the critical region (10) is expanding (see Fig. 1). Finally, at $\xi \simeq L_0$, both $\Delta_\xi$ and $E_c$ become of the order of $\Delta$, so that the WD regime shrinks to the quantum limit ($s \ll 1 \Leftrightarrow t \gg \hbar/\Delta$), the diffusive regime disappears entirely, and the critical regime expands to the whole [10] region $\omega \gtrsim \Delta$ (Fig. 1). There the TLCF becomes completely universal:

$$R_{ME}(s) = -\frac{A_{d\beta}}{s^{2-\gamma}}, \quad \gamma \equiv 1 - \frac{1}{\nu d}. \quad (19)$$

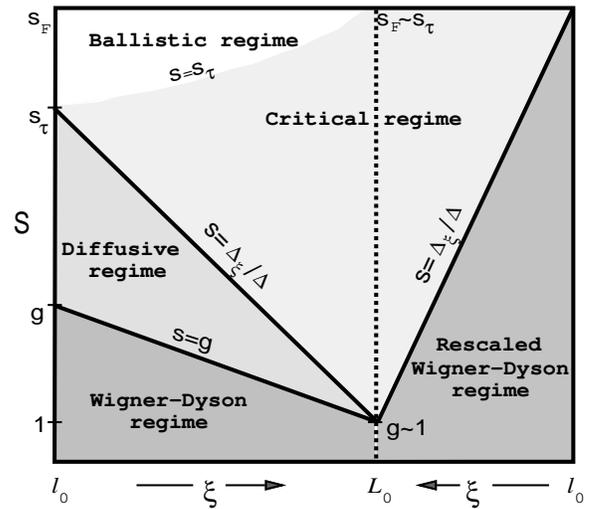

FIG. 1. Schematic phase diagram for the statistics in the crossover region from the metall (where the correlation length $\xi \lesssim \ell_o$) to the mobility edge (the dotted line, $\xi \gtrsim L_0$) to the insulator (where the localization length $\xi \lesssim \ell_o$). Rescaled WD regime is reduced to the WD regime at $\xi \gtrsim L_0$ and to the Poisson regime at $\xi/L_0 \to 0$, see Eq. (21). Here $s_\tau = \hbar/\tau\Delta$, and $s_F = \varepsilon_F/\Delta$.

This expression is not changed with a further increase of $\xi$, as for $\xi > L_0$ diffusion in the whole sample is anomalous and qualitatively the same as for $\xi \sim L_0$. Equation (19) coincides with that obtained directly at the mobility edge [8] within the diagrammatic approach.

In the above procedure, it seemed to be straightforward to estimate the numerical coefficient $\alpha$ in Eq. (17), and thus $A_{d\beta}$ in Eq. (18). However, we could find neither its value, nor even its sign, as this procedure was rather oversimplified. There are two reasons for this. First, the semi-classical expression (4) has been shown in Ref. [11] to be equivalent to the two-diffuson (and two-cooperon) diagrams of Ref. [3]. However, at the mobility edge and thus also in the crossover region, there are infinitely many of relevant diagrams. All of them have been proved to give parametrically the same contributions at the mobility edge [8]. Then, to get a parametrically correct answer it is sufficient to consider just the simplest typical contribution that can be described in the semiclassical language of Eq. (4). Second, we used the Gaussian form, Eq. (5), of the probability distribution $P(r, t)$, albeit with the time-dependent diffusion coefficient $D(t)$. In fact, in the region of anomalous diffusion, an exact expression for $P$ is not known. It remains causal, however. This, together with the scaling representation of its Fourier transform at the mobility edge, $\widetilde{P}(q, \varepsilon) \propto [F(q\xi, \Lambda_\varepsilon/\xi)q^d - i\varepsilon]^{-1}$ (with $\Lambda_\varepsilon \sim (\rho\varepsilon)^{-1/d}$ being a typical diffusion length during the time $\hbar/\varepsilon$ and $F$ being an *arbitrary* scaling function), allowed to prove [8] that the TLCF, Eq. (19), remains unchanged, except for the numerical coefficient. Such a proof that involves all possible diagrams is extended straightforwardly to the crossover region, as will be published elsewhere.

Finally, let us describe the statistics in the crossover region, $\ell_0 \ll \xi \lesssim L_0$, on the insulating side of the transition, where $\xi$ is the localization length. The level correlations are only due to the states confined to the same localization volume. The TLCF is obtained then as a superposition of independent contributions, $R_\xi$, from each volume $\xi^d$:

$$R_{\mathrm{ins}}(s) \simeq \delta(s) + \frac{\Delta}{\Delta_\xi} R_\xi\left(s\frac{\Delta}{\Delta_\xi}\right), \qquad (20)$$

where the $\delta(s)$ function describes the self-correlation of the levels.

At smaller energies, $\varepsilon < \Delta_\xi$, the argument of $R_\xi$ is smaller than 1, so that it should be qualitatively the same as the Wigner–Dyson TLCF which, in turn, becomes almost a constant for $s \ll 1$. Thus, we obtain in the region $\varepsilon \ll \Delta_\xi$:

$$R_{\mathrm{ins}}(s) \simeq \delta(s) + \frac{\Delta}{\Delta_\xi} R_{\mathrm{WD}}\left(s\frac{\Delta}{\Delta_\xi}\right) \approx \delta(s) - \frac{\Delta}{\Delta_\xi}. \qquad (21)$$

The regular part is small as only the small fraction of the levels, $(\xi/L_0)^d = \Delta/\Delta_\xi$, are correlated. The Poisson statistics emerge in the thermodynamic limit, where this regular tail vanishes.

At larger energies, we arrive at the critical regime of Eq. (10), where the scaling relations differ from those on the metallic side, Eqs. (15) and (16), only by the sign of the $\xi$ dependent corrections. Thus, the TLCF on the insulating side may differ from that on the metallic side, Eq. (18), only by a numerical coefficient. Although this coefficient is unknown, as explained above, it should be equal to that in Eq. (19). Indeed, there is no distinction between metal and insulator at the critical point where $\Delta = \Delta_\xi$, and Eq. (20) should coincide for $s \gg 1$ with Eq. (19), obtained at the mobility edge. Then one should substitute $R_{\mathrm{ME}}$ for $R_\xi$ into Eq. (20), which then becomes identical to Eq. (18), obtained in the critical regime (10) on the metallic side. So, the critical statistics, universal at the mobility edge, are equally applicable on both sides of the metal-insulator transition in the crossover region.

**ACKNOWLEDGMENTS**

We are thankful to B. L. Altshuler, Y. Gefen, V. V. Lebedev, A. D. Mirlin, and B. Shapiro for usefull discussions. A.G.A. is grateful to GIF for financial support and ICTP for kind hospitality. V.E.K. and I.V.L. gratefully acknowledge travelling support under the EEC grant No. SSC-CT90-0020.